\documentclass{article}

 \usepackage[sglblindworkshop, final]{neurips_2025}
\workshoptitle{2nd Workshop on Multi-modal Foundation Models and Large Language Models for Life Sciences}

\usepackage[utf8]{inputenc} 
\usepackage[T1]{fontenc}    
\usepackage{hyperref}       
\usepackage{url}            
\usepackage{booktabs}       
\usepackage{amsfonts}       
\usepackage{nicefrac}       
\usepackage{microtype}      
\usepackage{xcolor}         
\usepackage{amsmath}

\usepackage{subcaption}
\usepackage{graphicx}
\usepackage{enumitem}
\usepackage{bm}

\usepackage[table]{xcolor}
\definecolor{rowgray}{gray}{0.92} 

\newcommand{\beginsupplement}{%
        \setcounter{table}{0}
        \renewcommand{\thetable}{S\arabic{table}}%
        
        \setcounter{figure}{0}
        \renewcommand{\thefigure}{S\arabic{figure}}%
    
        \setcounter{equation}{0}
        \renewcommand{\theequation}{S\arabic{equation}}%
        
        \setcounter{section}{0}
        \renewcommand{\thesection}{S\arabic{section}}%
        
        \setcounter{page}{1}
        \renewcommand{\thepage}{S\arabic{page}}%
}

\title{Quantifying the Role of OpenFold Components in Protein Structure Prediction}

\author{%
  Tyler~L.~Hayes \\
  College of Computing\\
  Georgia Institute of Technology\\
  \texttt{thayes68@gatech.edu} \\
  \And
  Giri~P.~Krishnan \\
  College of Computing\\
  Georgia Institute of Technology\\
  \texttt{gkrishnan46@gatech.edu} \\
}

\begin{document}

\maketitle

\begin{abstract}
    Models such as AlphaFold2 and OpenFold have transformed protein structure prediction, yet their inner workings remain poorly understood. We present a methodology to systematically evaluate the contribution of individual OpenFold components to structure prediction accuracy. We identify several components that are critical for most proteins, while others vary in importance across proteins. We further show that the contribution of several components is correlated with protein length. These findings provide insight into how OpenFold achieves accurate predictions and highlight directions for interpreting protein prediction networks more broadly.
\end{abstract}

\section{Introduction}\label{intro}

Recent protein structure prediction models such as AlphaFold2~\cite{jumper2021af2} and OpenFold~\cite{ahdritz2024openfold} have transformed biology, enabling breakthroughs in protein folding~\cite{varadi2024alphafold}, drug discovery~\cite{karelina2023accurately,kovalevskiy2024alphafold}, and protein synthesis~\cite{goverde2023novo,varadi2024alphafold}. Yet, despite their success, it remains unclear how these models achieve such accuracy, i.e., which architectural components are most essential for prediction. At the core of several protein structure prediction models lies a transformer-based model that iteratively refines two key internal representations: the multiple sequence alignment (MSA) representations and the pairwise residue (Pair) representations. In AlphaFold2 and OpenFold, the main transformer model is called the Evoformer, which includes several components such as attention layers, transition MLPs, and triangular update operations. However, the relative importance of these components for structure prediction is not well understood. Previous ablations of AlphaFold2 and OpenFold focused mainly on auxiliary losses, training regimes, or coarse architectural changes (e.g., ``no triangles, biasing, or gating''), leaving the role of individual Evoformer components largely unexplored. In this paper, we address this gap by systematically evaluating component-level contributions across proteins. Our study reveals which components are broadly critical, which are dispensable, and how their importance varies with properties such as protein length.

\textbf{Our main contributions are as follows:}
\begin{enumerate}
    \item We propose a methodology to systematically examine the contribution of individual OpenFold components to structure prediction accuracy, revealing protein-specific reliance on different architectural components.
    \item We identify several components that are critical for accurate predictions across most proteins, including MSA Column Attention, both MLP Transition layers, and the final Pair representation used for structure prediction, providing biological insights into \emph{how} OpenFold achieves accurate predictions.
    \item We analyze how these contributions vary with protein length and find that the importance of several components is statistically significantly correlated with sequence length.
\end{enumerate}

\section{Related Work}\label{related-work}

Several recent works have studied the inner workings of AlphaFold2. For example, \cite{tan2023explainablefold} proposes an objective function to solve for a set of residue deletions or substitutions required to change the network's structure predictions. In \cite{gut2025dissecting}, the impact of different templates and the recycling mechanism are studied in more detail. In \cite{sibli2025enhancing}, a framework is proposed to study the contribution of different amino acids to final structure prediction using SHAP. Our study differs from existing studies since we seek to characterize the contribution of architectural components to structure prediction, rather than the contribution of individual residues, templates, or recycling iterations. Beyond structure prediction models, there has also been interpretability work on protein language models (see \cite{nguyen2025advances} for a survey). In \cite{garcia2025interpreting,simon2024interplm}, sparse auto-encoders are used to study the ESM-2~\cite{lin2023evolutionary} representation space. There has also been work to study the intrinsic dimensionality of protein language model representation spaces~\cite{facco2019intrinsic,valeriani2023geometry,yang2024probing}. In \cite{vig2021bertology}, a methodology is proposed to relate the attention maps of protein language models to protein properties such as structure or function. More broadly, the interpretability of protein models has been proposed as a promising future direction by several researchers~\cite{chen2023protein,chen2024applying,kovalevskiy2024alphafold,nguyen2025advances}.

\section{Methods}\label{methods}

\paragraph{Model Components.}

\begin{figure}[t]
  \centering
  \includegraphics[width=0.80\linewidth]{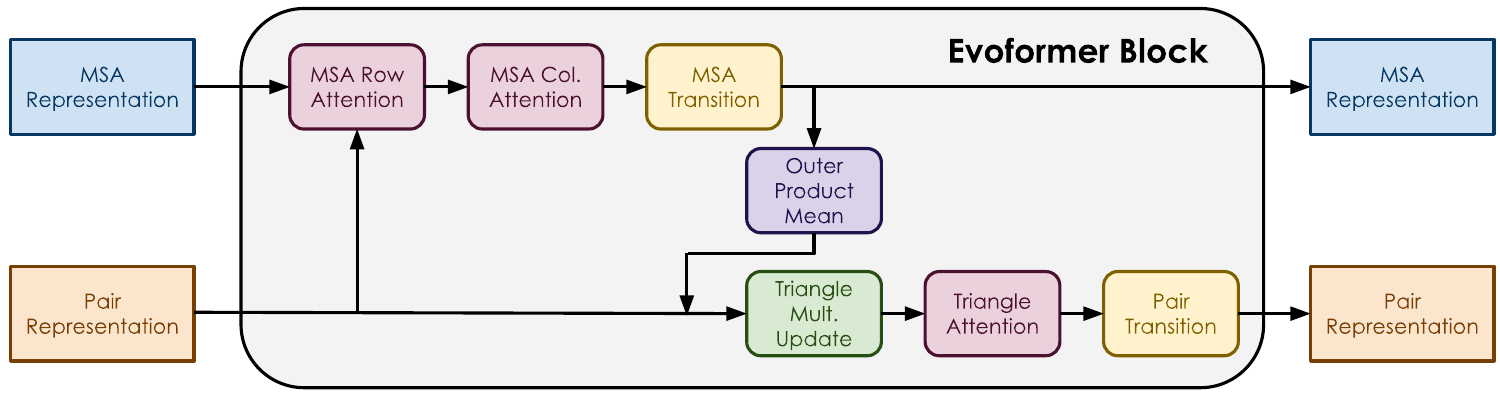}
  \caption{\textbf{One Evoformer Block in OpenFold.} Each block operates on the MSA and Pair representations via a series of attention, transition, and update operations. For clarity, residual connections are omitted, except for the Outer Product Mean connection from the MSA to the Pair representation.}
  \label{fig:evoformer}
\end{figure}

OpenFold~\cite{ahdritz2024openfold} is an open-source reproduction of AlphaFold2~\cite{jumper2021af2}, making it a suitable model for our study. Despite AlphaFold2’s advances in protein structure prediction~\cite{rives2021biological}, many architectural contributions remain unclear~\cite{kovalevskiy2024alphafold}. Since recent models such as AlphaFold3~\cite{abramson2024af3} and Boltz~\cite{passaro2025boltz,wohlwend2024boltz} retain the same transformer-based architecture with triangular operations on Pair representations, our findings extend beyond OpenFold. OpenFold operates in three distinct phases to predict structures from amino acid sequences: i) a pre-processing phase that produces a Multiple Sequence Alignment (MSA) representation and a Pair representation; ii) Evoformer processing via 48 blocks to refine these representations; and iii) the Structure Module, which outputs a 3D structure from these representations. The MSA representation is produced by comparing the input sequence to existing sequences from nature, while the Pair representation is produced by comparing residue-residue pair relationships in the sequence.

Here, we study the impact of various Evoformer components on structural accuracy (see Fig.~\ref{fig:evoformer} for an overview). Specifically, each Evoformer block consists of two pathways to operate on the MSA and Pair representations. Within the MSA pathway, MSA Row Attention integrates information across homologous sequences for each residue position, followed by MSA Column Attention, which correlates residues along each sequence. An MSA Transition (feedforward MLP) further transforms these features. The Outer Product Mean projects the MSA representation into the Pair representation, which is refined through Triangle Multiplicative Updates and Triangle Attention, designed to enforce the triangle inequality for geometric consistency among residue triplets. Finally, the Pair Transition applies a feedforward MLP before the final representations are passed to the next Evoformer block, or the Structure Module. There are also residual connections before and after each module.

\paragraph{Analysis Procedure.}

We perform three types of experiments: i) skipping attention modules, ii) skipping non-attention modules or zeroing out representations, and iii) evaluating the correlation between the change in performance due to component manipulation with (protein) sequence length. Skipping is implemented by bypassing the module’s operations in all Evoformer blocks, while zeroing is implemented by setting all values of the final representation to zero before passing it to the Structure Module. Component-level experiments identify which modules affect structure prediction the most, while representation-level experiments assess the usefulness of each representation. For length analyses, we plot the difference in TM-score between the baseline and component removal experiment as a function of protein length. We then fit a line to the data via linear regression and perform Spearman correlation analyses.

\paragraph{Data \& Metrics.}

We use the three-month CAMEO~\cite{haas2018continuous} subset from OpenFold, restricted to proteins with fewer than 700 residues~\cite{ahdritz2024openfold}. After filtering out targets with missing structural files and runs with baseline TM-scores below 0.7, we obtain 154 proteins. We run OpenFold \texttt{model\_1\_ptm} with the AlphaFold2 JAX weights and report results with unrelaxed structures. Our primary metric is the TM-score between predictions and experimental ground truth. To quantify the effect of each component, we compare the baseline TM-score to the score when a component is skipped or zeroed ($\Delta$ TM). For length correlation analyses, we report $R^2$, Spearman’s $\rho$, and $p$-values. We run experiments with each protein three times and use the average performance across runs for our analyses. More details are in Supplemental Materials (Sec.~\ref{sec:data}), including raw TM-scores (Fig.~\ref{fig:violin-full-raw}).

\section{Results: Component-Level Findings}\label{results}

\begin{figure}[t]
  \centering
  \begin{subfigure}[t]{0.46\linewidth}
    \centering
    \includegraphics[width=\linewidth]{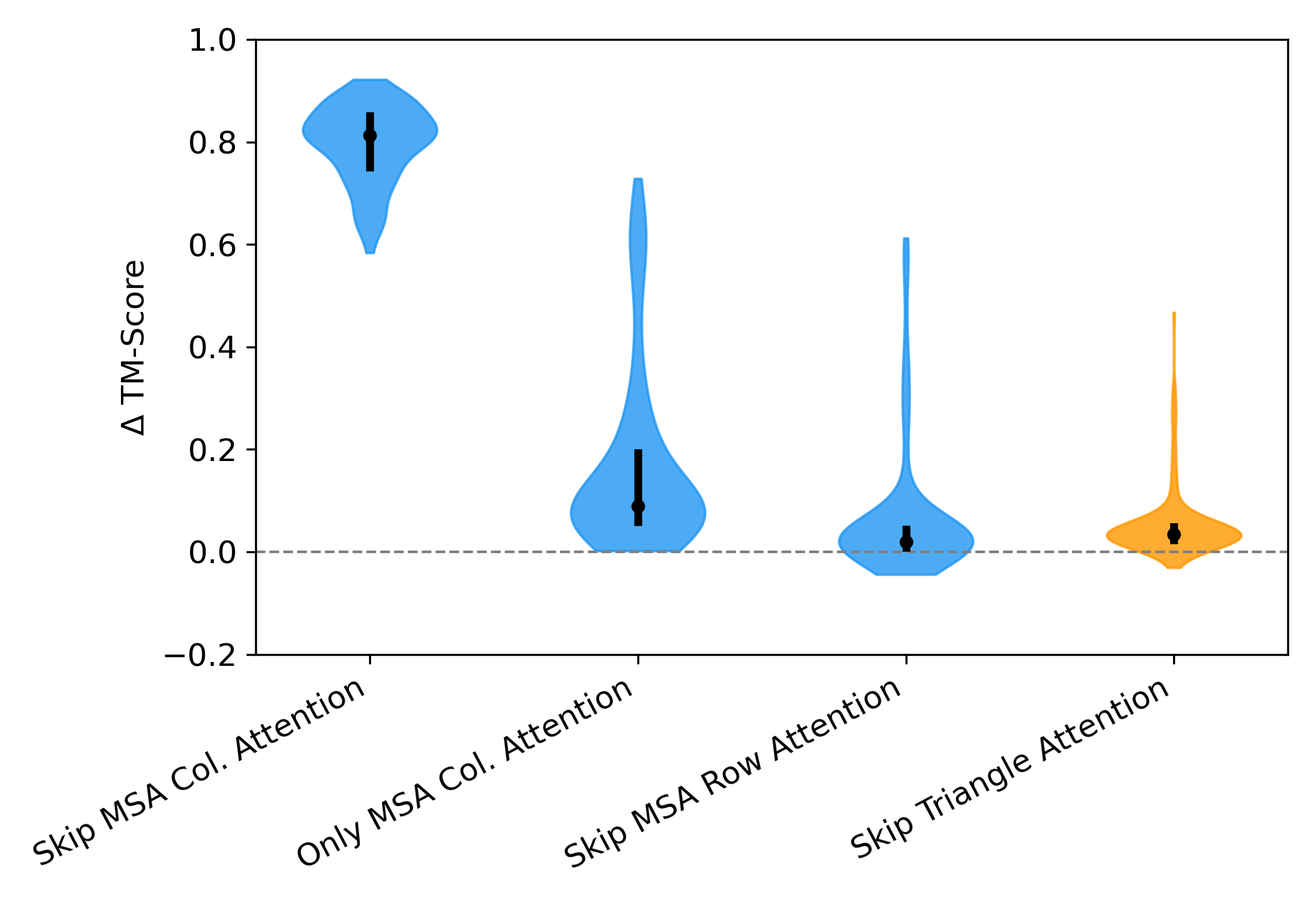}
    \caption{Attention Experiments}
    \label{fig:attention}
  \end{subfigure}\hfill
  \begin{subfigure}[t]{0.46\linewidth}
    \centering
    \includegraphics[width=\linewidth]{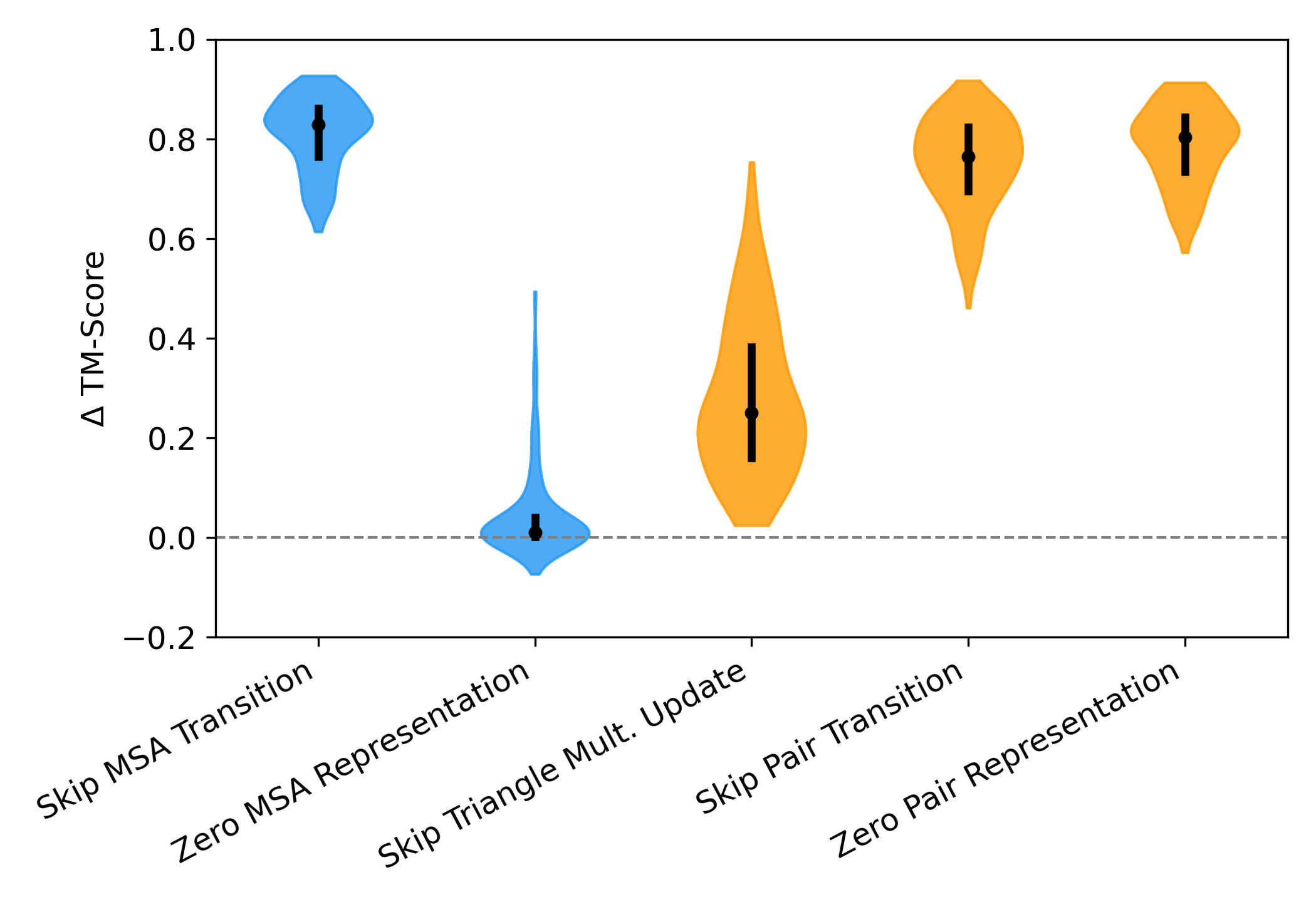}
    \caption{Non-Attention Experiments}
    \label{fig:non-attention}
  \end{subfigure}\
  \caption{\textbf{Differences ($\Delta$) Between Baseline TM-Scores and Component Studies Across Proteins.} 
  Higher scores indicate more deviation from the Baseline. Studies involving MSA and Pair Representations are in \color{blue}{blue} \color{black}{and} \color{orange}{orange}\color{black}{, respectively.}
  }
  \label{fig:violin}
\end{figure}


\begin{figure}[t]
  \centering
  \begin{subfigure}[t]{0.32\linewidth}
    \centering
    \includegraphics[width=\linewidth]{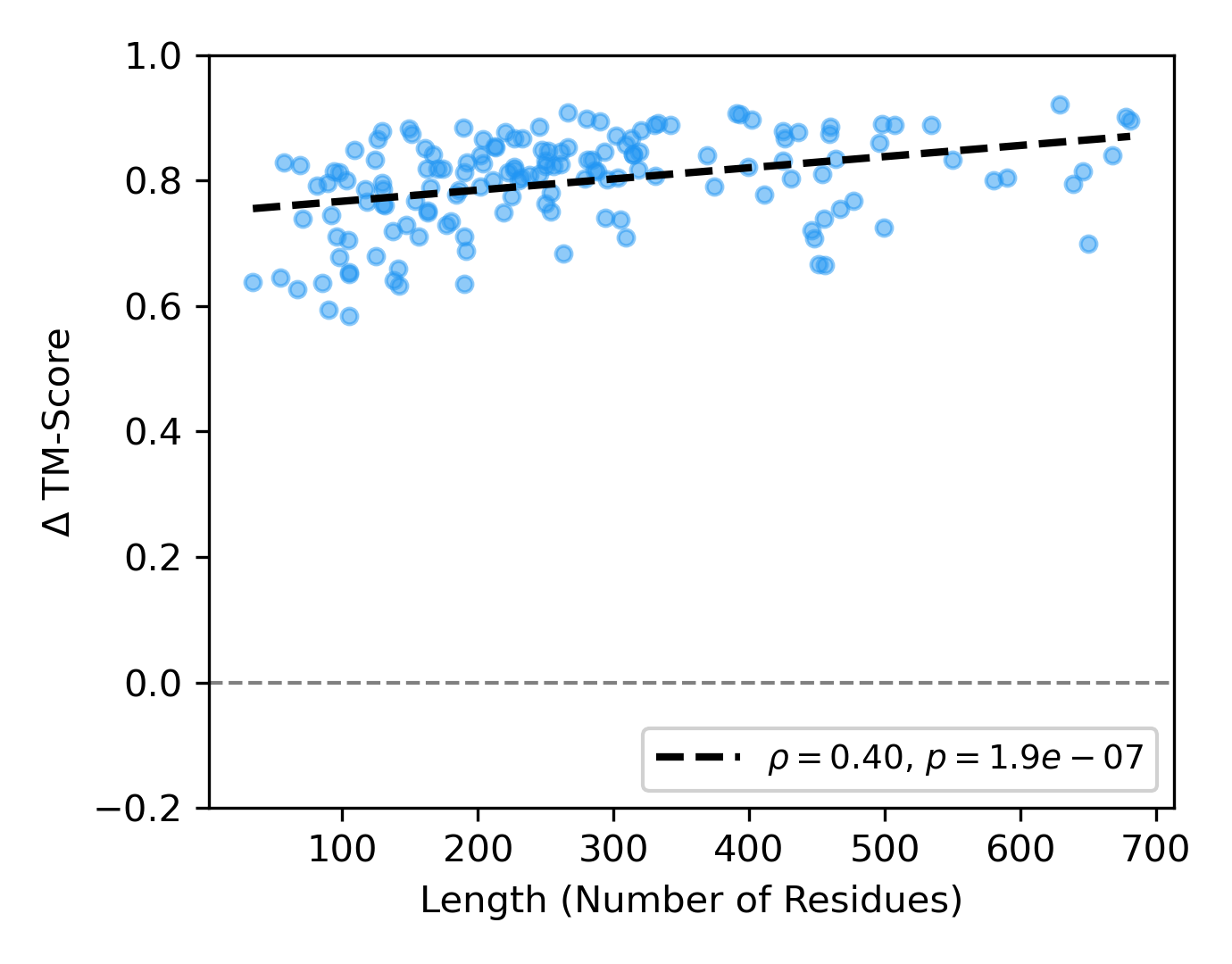}
    \caption{Skip MSA Col. Attention}
    \label{fig:del-msa-col}
  \end{subfigure}\hfill
    \begin{subfigure}[t]{0.32\linewidth}
    \centering
    \includegraphics[width=\linewidth]{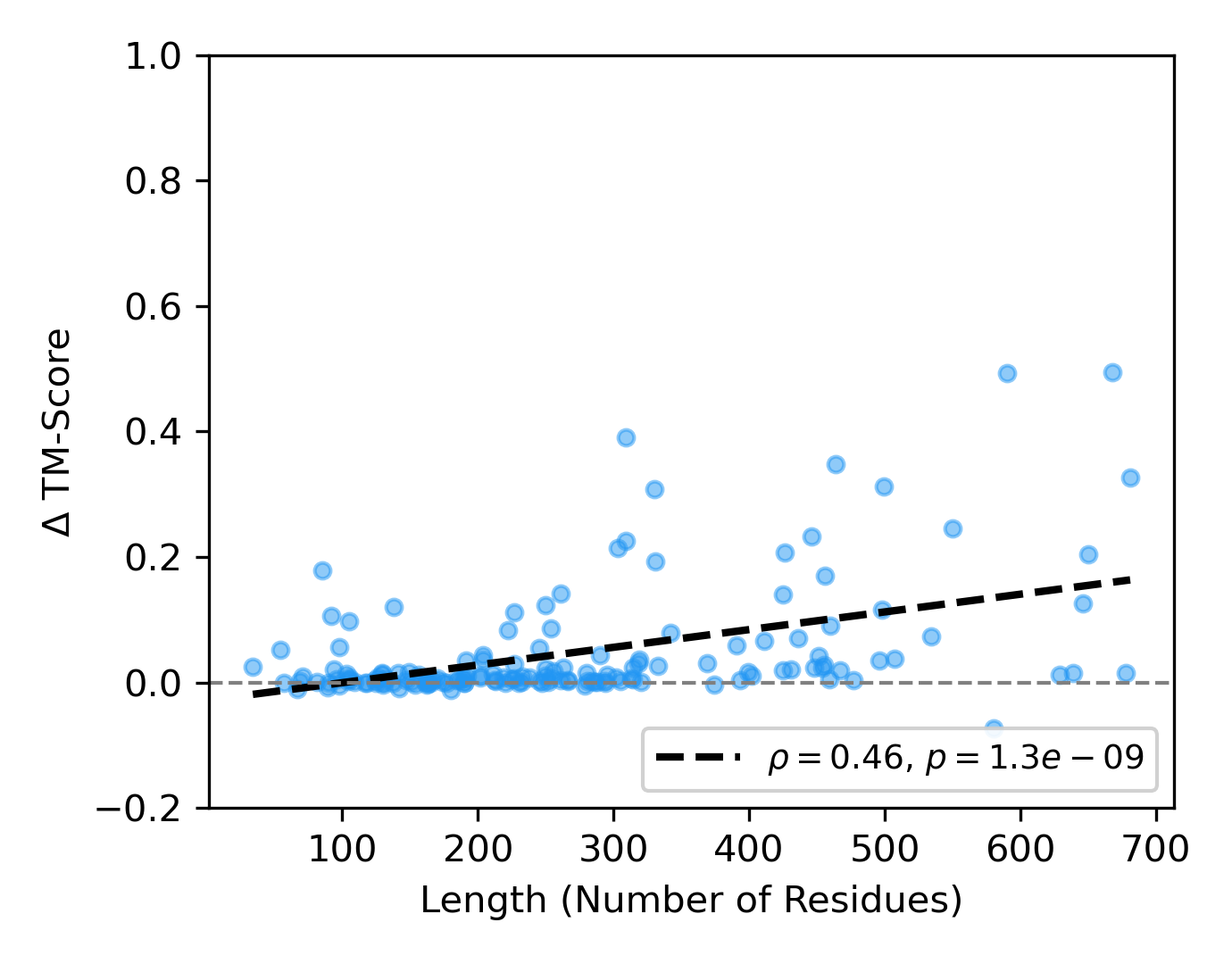}
    \caption{Zero MSA Representation}
    \label{fig:zero-msa}
  \end{subfigure}\hfill
  \begin{subfigure}[t]{0.32\linewidth}
    \centering
    \includegraphics[width=\linewidth]{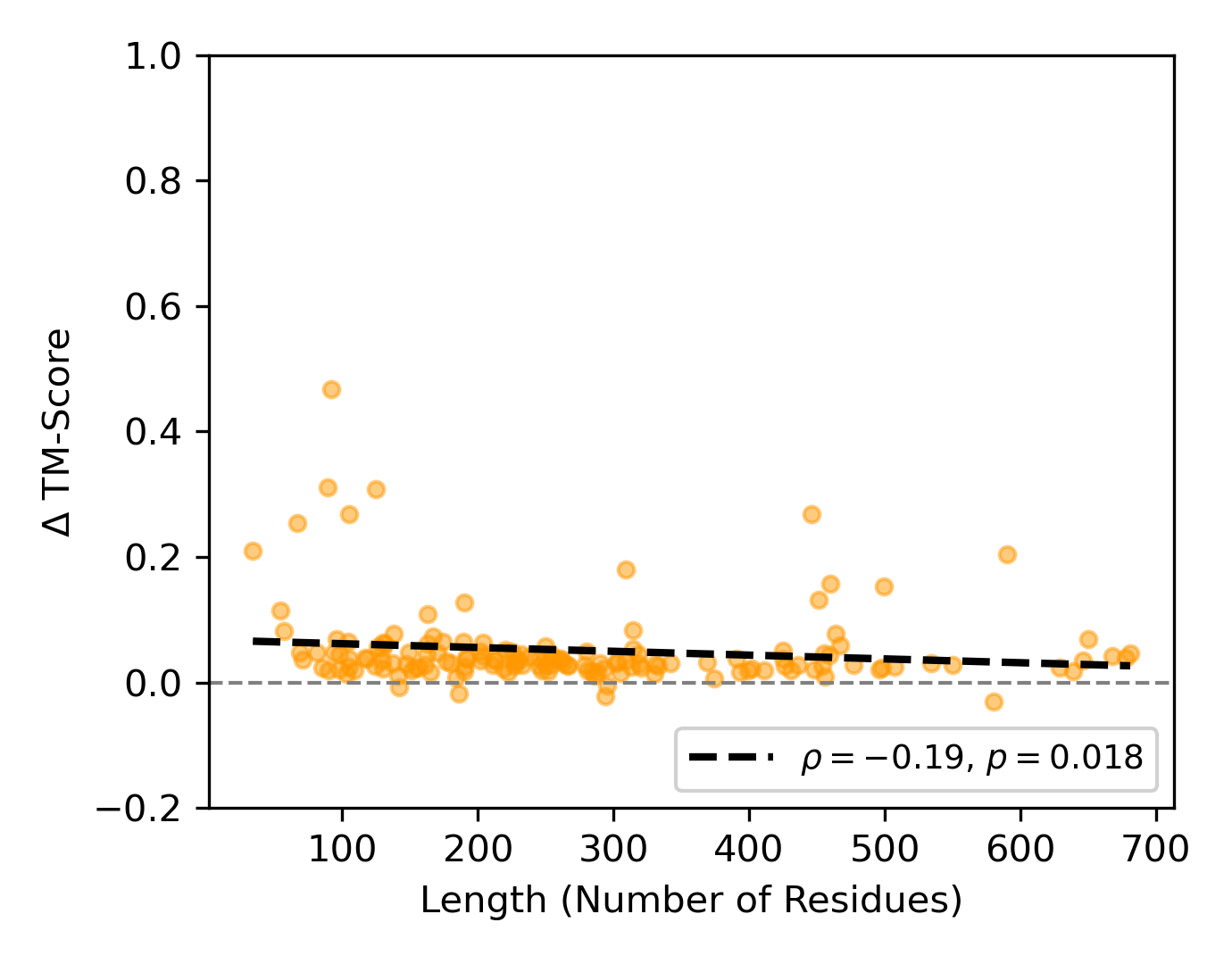}
    \caption{Skip Triangle Attention}
    \label{fig:del-tri}
  \end{subfigure}\hfill
  \caption{\textbf{Comparison of Baseline Differences in TM-Scores ($\Delta$) vs. Protein Length.}
  Higher scores indicate more deviation from the Baseline. We show the best-fit line, the Spearman correlation coefficient, and its associated $p$-value. Each point represents the results for one protein.}
  \label{fig:scatter-plots}
\end{figure}

\paragraph{Analyzing Attention Components.}

In Fig.~\ref{fig:attention}, we study the impact of skipping MSA Column Attention, MSA Row Attention, and Triangle Attention (around the starting and ending nodes) in the Evoformer. Most proteins are negatively impacted by skipping MSA Column Attention, the largest deviation from baseline. In contrast, bypassing MSA Row Attention has minor performance impacts for most proteins, and bypassing Triangle Attention has minimal impacts on most proteins. However, the broad variance of the violin plots for these experiments indicate that some proteins benefit from the inclusion of MSA Row or Triangle attention. Since MSA Column Attention was most critical, we also tested models with \emph{only} this attention component, i.e., skip MSA Row and Triangle attention. Surprisingly, we found that many proteins performed well with only MSA Column Attention, showing only slight deviations from the baseline (median $\Delta$TM is 0.089). This implies that OpenFold heavily relies on evolutionary relationships for predicting structures. We also study the impact of using only MSA Row Attention or only Triangle Attention (Fig.~\ref{fig:attention-full}) and find that these attention types are not enough on their own for structure prediction.

\paragraph{Analyzing Non-Attention Components.}

In Fig.~\ref{fig:non-attention}, we evaluate the impact of the MSA Transition, the Pair Transition, the Triangle Multiplicative Update (outgoing and incoming), and the MSA and Pair representations on structure prediction. Zeroing the MSA representation had the smallest effect, while skipping transitions or zeroing the Pair representation caused the largest drops. The high variance of the violin plot for zeroing out the MSA representation indicates that some proteins benefit from the inclusion of this component. Removing transition layers results in large performance drops (median $\Delta$TM is 0.765 and 0.829 for skipping Pair and MSA transitions, respectively), consistent with previous work suggesting that MLP layers in transformers contain crucial semantics~\cite{he2024matters,lin2024mlp}. Skipping the Triangle Multiplicative Update produced highly variable outcomes across proteins, in contrast to Triangle Attention, which had little effect. This suggests that the update is more critical for performance. Removing both operations (Fig.~\ref{fig:non-attention-full}) led to poor accuracy, underscoring the heterogeneity in which components OpenFold relies on.

\paragraph{Correlating Component Removal with Protein Length.}

In Fig.\ref{fig:scatter-plots}, we plot $\Delta$TM-score versus protein length for three key experiments: skipping MSA Column Attention, zeroing the MSA representation, and skipping Triangle Attention (others are in Fig.\ref{fig:scatter-plots-supp}). Summary statistics ($R^2$, Spearman’s $\rho$, $p$-value) are in Table~\ref{tab:correlation}. Skipping MSA Column Attention led to increasing performance loss with length ($p=1.9 \times 10^{-7}$), indicating stronger reliance in longer proteins. Zeroing the MSA representation showed little effect on short proteins but larger losses for longer proteins ($p=1.3 \times 10^{-9}$). In contrast, skipping Triangle Attention correlated negatively with length ($p=0.018$), suggesting greater reliance on Triangle Attention for predicting structures for shorter proteins. Other components (Pair representation, MSA/Pair Transitions) also showed length dependence, while skipping MSA Row Attention, only using MSA Column Attention, and skipping the Triangle Multiplicative Update did not. This suggests that length only partially explains the heterogeneity in component importance, and additional factors such as fold type could be studied in the future.

\begin{table}[t]
\caption{\textbf{Correlation Between $\Delta$TM and Protein Length.} We report $R^2$, Spearman’s $\rho$, and $p$-values for each experiment. We show $p<0.05$ in \textbf{bold}.
\label{tab:correlation}}
\centering
\resizebox{\linewidth}{!}{
\begin{tabular}{lccclccc}
\toprule
\textsc{Attention Study} & $R^2$ & $\rho$ & $p$-value & \textsc{Non-Attention Study} & $R^2$ & $\rho$ & $p$-value \\ 
\midrule
Skip MSA Col. Attention & 0.13 & 0.40 & \textbf{1.9e-7} & Skip MSA Transition & 0.09 & 0.34 & \textbf{1.2e-5} \\
Only MSA Col. Attention & 0.02 & -0.13 & 0.11 & Zero MSA Representation & 0.21 & 0.46 & \textbf{1.3e-9} \\
Skip MSA Row Attention & 0.01 & -0.07 & 0.42 & Skip Triangle Mult. Update & 0.06 & 0.08 & 0.31 \\
Skip Triangle Attention & 0.02 & -0.19 & \textbf{0.018} & Skip Pair Transition & 0.26 & 0.56 & \textbf{3.8e-14} \\
 & & & & Zero Pair Representation & 0.11 & 0.38 & \textbf{1.1e-6} \\
\bottomrule
\end{tabular}
}
\end{table}

\section{Discussion \& Conclusion}\label{conclusion}

We showed that OpenFold’s structure prediction relies most heavily on MSA Column Attention, the MLP Transition layers, and the final Pair representation. Other components, including MSA Row Attention, Triangle Attention, the Triangle Multiplicative Update, and the final MSA representation, contributed variably across proteins, with the multiplicative update showing the widest variance. Remarkably, much of the baseline performance could be recovered using only MSA Column Attention, highlighting the reliance on evolutionary sequence information. Several components, including MSA Column and Triangle Attention, the Transition layers, and the MSA/Pair representations, exhibited length-dependent effects: longer proteins depended more on MSA-based features, whereas shorter proteins were more sensitive to Triangle Attention. In contrast, the Triangle Multiplicative Update showed no length dependence, indicating heterogeneity beyond sequence length. Overall, different proteins rely on different Evoformer components. Future work could test additional factors, such as fold type or function, to further explain this heterogeneity. Our findings provide key insights into the inner workings of AlphaFold-like models.

\section*{Acknowledgements}

This work is supported in part by NSF award \#2502793. This research is also supported by the National Artificial Intelligence Research Resource (NAIRR) Pilot and the Delta advanced computing and data resource, which is supported by the National Science Foundation (award NSF-OAC \#2005572). The views and conclusions contained herein are those of the authors and should not be interpreted as representing the official policies or endorsements of any sponsor.



{
\small
\bibliographystyle{plain}
\bibliography{sn-bibliography}
}

\newpage
\appendix
\beginsupplement

\section{Supplemental Materials}

\section{Data \& Metrics} \label{sec:data}

We follow the OpenFold paper~\cite{ahdritz2024openfold} and use a three-month subset of the CAMEO dataset (ending on January 16, 2022) of proteins with fewer than 700 residues~\cite{haas2018continuous}. This subset contains 185 proteins, of which we exclude one protein (7eqx\_C) due to the removal of its mmCIF file (4uf4) from the RCSB website. This CAMEO subset is ideal for validating the impact of components on model performance as it was not used during training. We run OpenFold \texttt{model\_1\_ptm} with the suggested paper settings and the original AlphaFold2 JAX weights\footnote{\url{https://github.com/aqlaboratory/openfold}}. While three recycles are typically used, we found minor performance differences between using zero and three recycles (0.7\% on the CAMEO subset), which is consistent with previous findings~\cite{gut2025dissecting}, so we use zero recycles here. Additionally, we use the unrelaxed structure predictions from the model, which have minimal performance differences with the relaxed predictions (0.003\% difference). Our main performance metric is the TM-score between model predictions and experimental ground truth from the RCSB website. For all experiments, we filter proteins down to those that achieve a TM-Score of greater than 0.7 with a standard OpenFold run. This yields a set of 154 proteins and eliminates poor performance as a factor that could contribute to our component studies. To capture the influence of each component, regardless of model performance on specific proteins, we show the differences ($\Delta$) between a Baseline run of OpenFold (without any components skipped) and the study with the component skipped or zeroed out. For our studies correlating performance with protein length, we report $R^2$ of the best-fit line, the Spearman correlation coefficient ($\rho$), and the Spearman $p$-value.

\section{Raw TM-Score Plots}

\begin{figure}[h]
  \centering
  \begin{subfigure}[t]{0.51\linewidth}
    \centering
    \includegraphics[width=\linewidth]{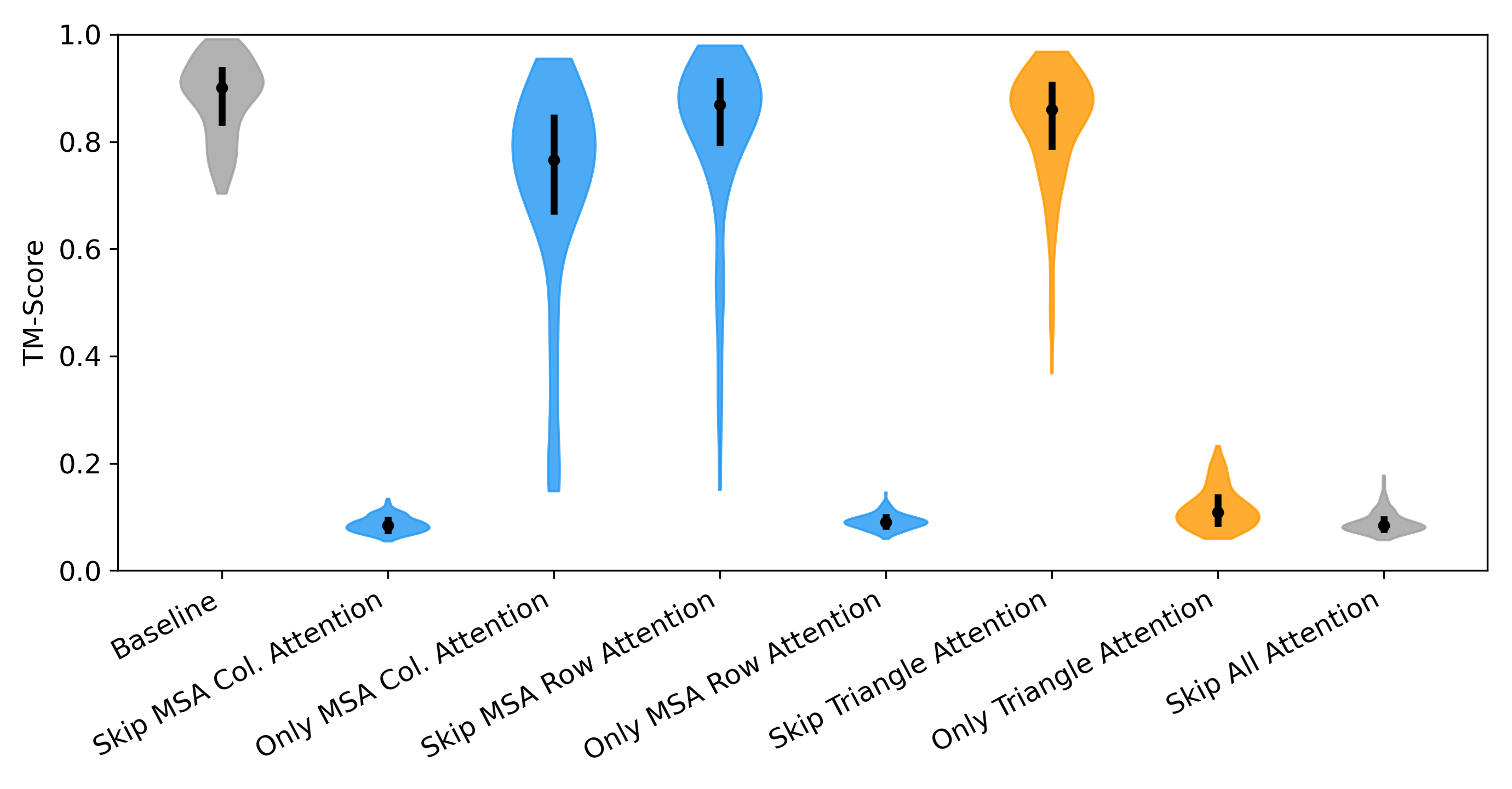}
    \caption{Attention Experiments}
    \label{fig:attention-full-raw}
  \end{subfigure}\hfill
  \begin{subfigure}[t]{0.46\linewidth}
    \centering
    \includegraphics[width=\linewidth]{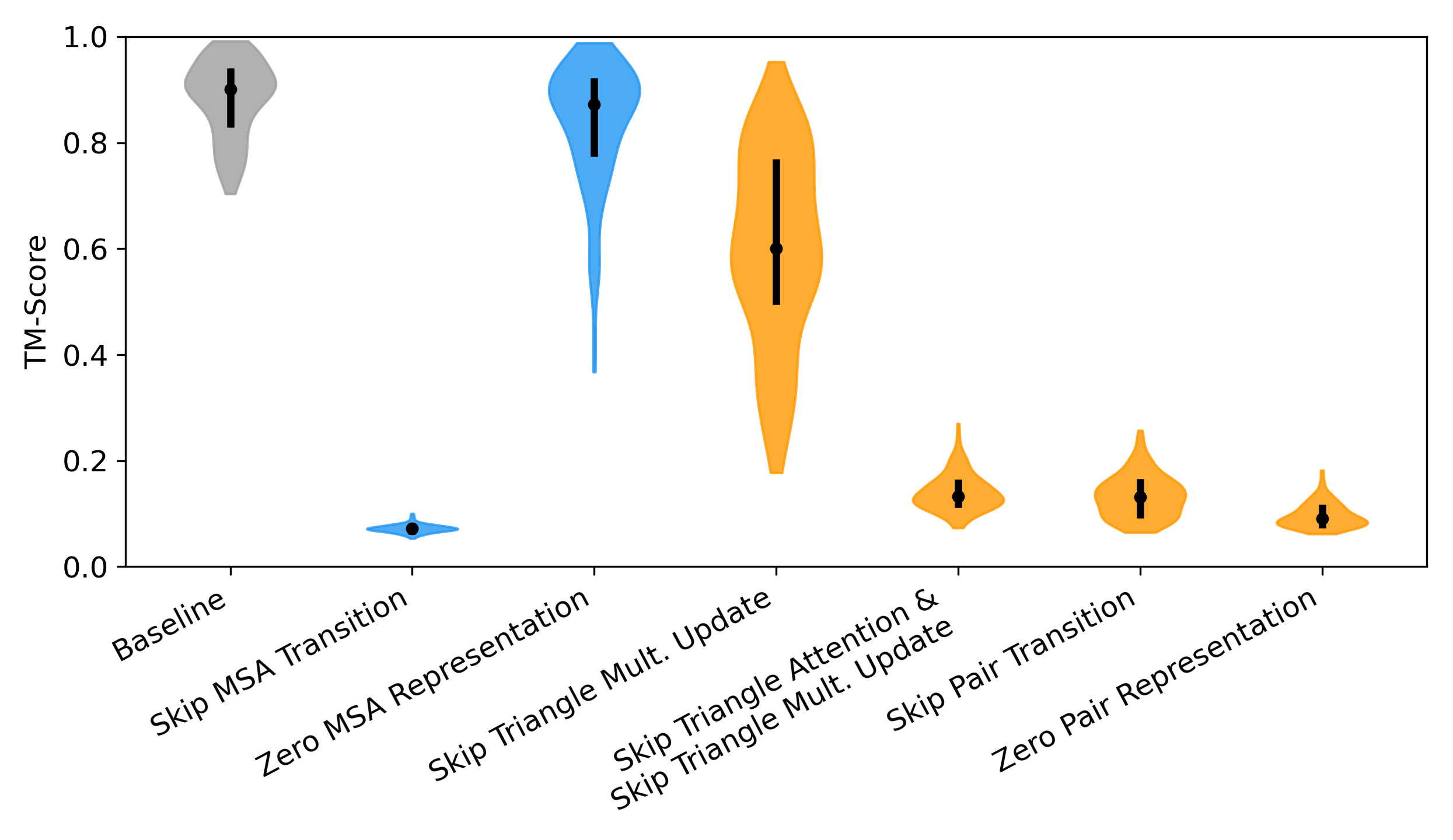}
    \caption{Non-Attention Experiments}
    \label{fig:non-attention-full-raw}
  \end{subfigure}\
  \caption{\textbf{Raw TM-Scores for Component Studies.} 
  Higher scores indicate better performance. Studies with MSA and Pair Representations are in \color{blue}{blue} \color{black}{and} \color{orange}{orange}\color{black}{, respectively.}
  }
  \label{fig:violin-full-raw}
\end{figure}

We show the raw TM-Scores for each component study in Fig.~\ref{fig:violin-full-raw}, including the Baseline performance. We use Baseline differences for plots in our main paper (Fig.~\ref{fig:violin}) to account for some protein structures being easier to predict than others. This allows us to focus directly on trends without model performance conflating the trends.

\section{Analyzing the Impact of More Components}

\begin{figure}[h]
  \centering
  \begin{subfigure}[t]{0.51\linewidth}
    \centering
    \includegraphics[width=\linewidth]{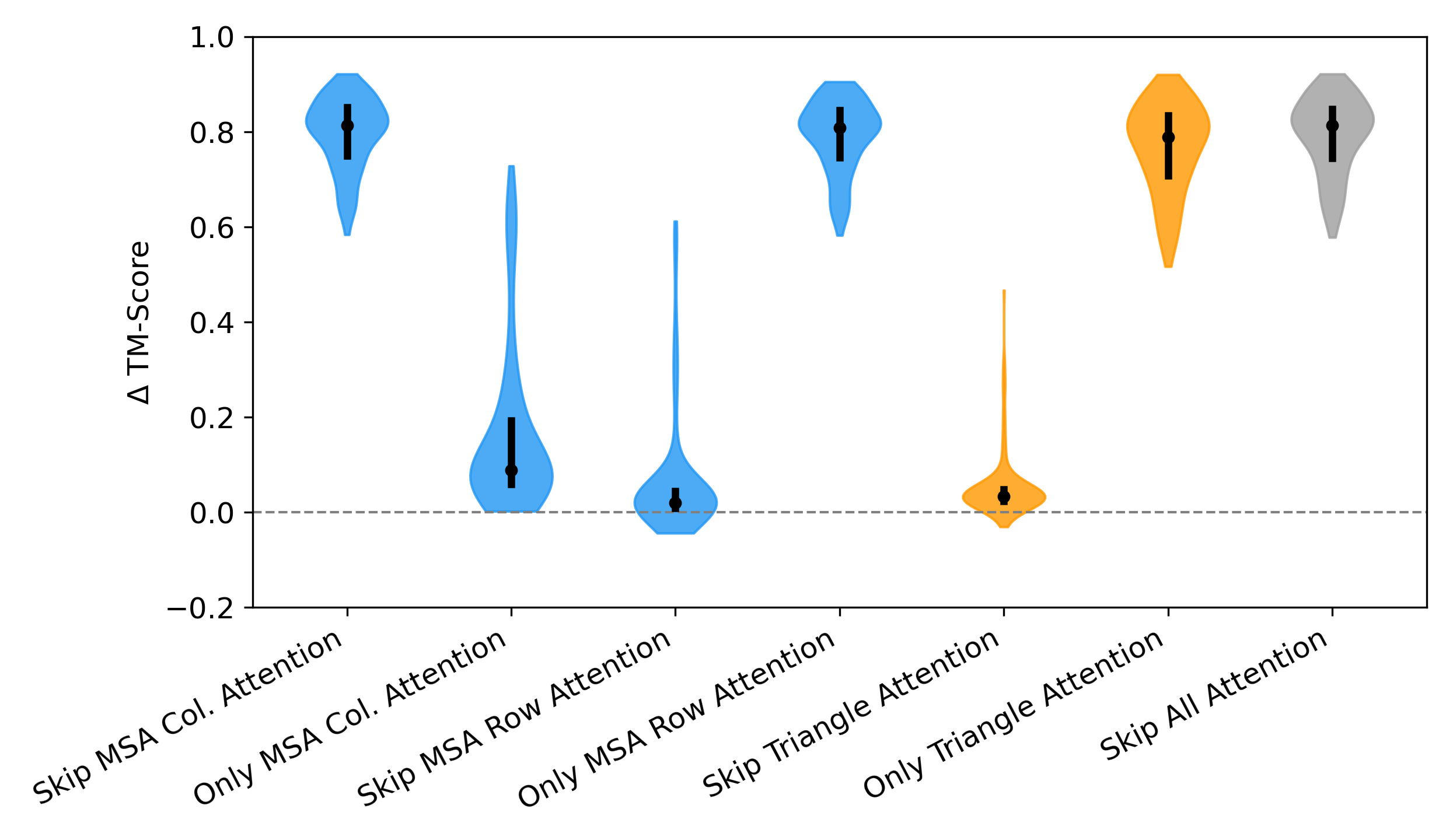}
    \caption{Attention Experiments}
    \label{fig:attention-full}
  \end{subfigure}\hfill
  \begin{subfigure}[t]{0.46\linewidth}
    \centering
    \includegraphics[width=\linewidth]{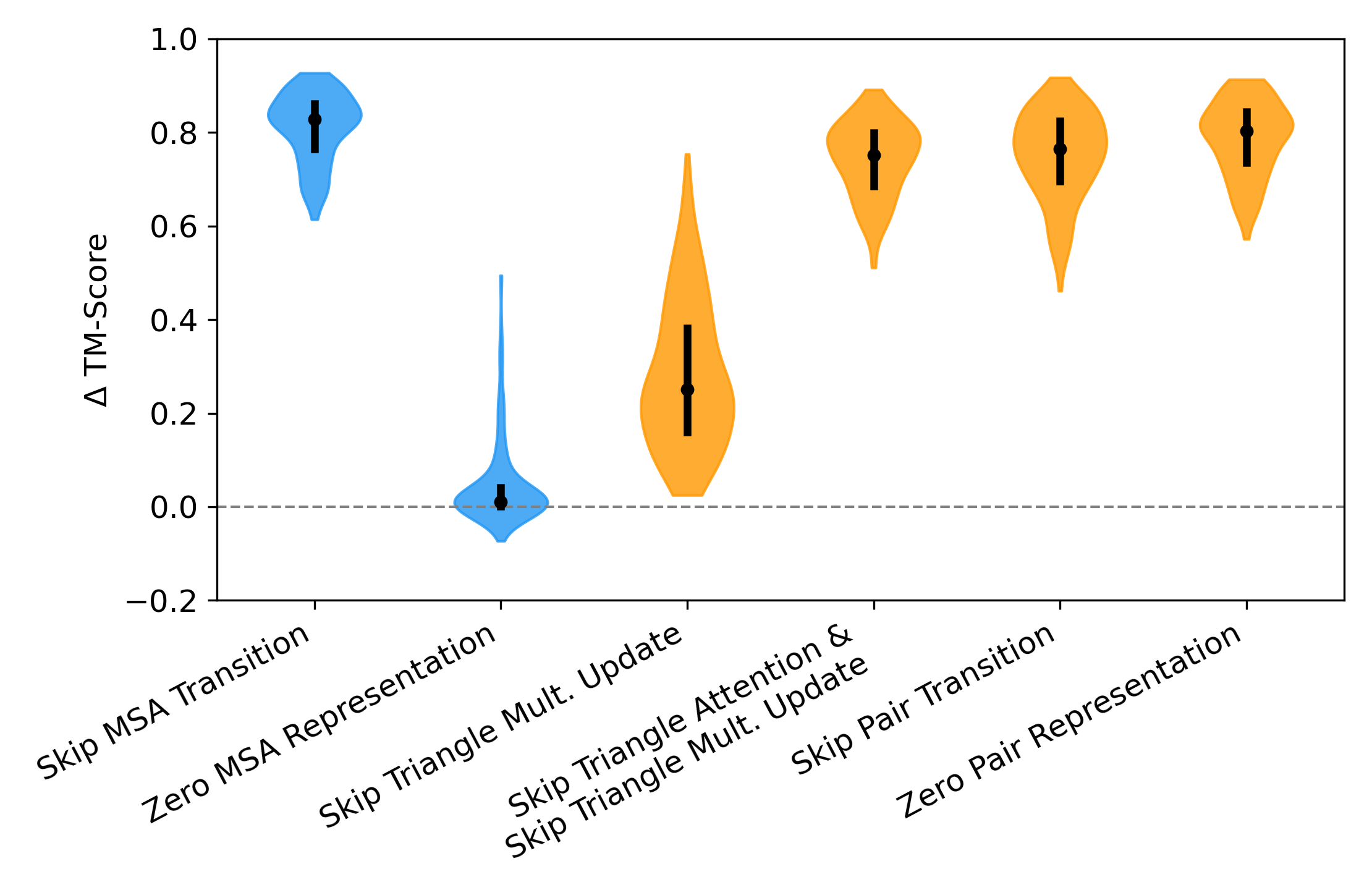}
    \caption{Non-Attention Experiments}
    \label{fig:non-attention-full}
  \end{subfigure}\
  \caption{\textbf{Differences ($\Delta$) Between Baseline TM-Scores and Component Studies Across Proteins.} 
  Higher scores indicate more deviation from the Baseline. Studies involving MSA and Pair Representations are in \color{blue}{blue} \color{black}{and} \color{orange}{orange}\color{black}{, respectively.}
  }
  \label{fig:violin-full}
\end{figure}

In Fig.~\ref{fig:violin-full}, we show the differences between the Baseline and various component studies for a larger set of experiments than our main paper. In Fig.~\ref{fig:attention-full}, we additionally show the impact of using only MSA Row Attention, only Triangle Attention, or skipping attention altogether. We find that only using MSA Row Attention or only Triangle Attention is not sufficient for protein structure prediction with OpenFold as there is a large difference in performance with the Baseline. Moreover, in Fig.~\ref{fig:non-attention-full}, we additionally show the impact of skipping both Triangle Attention and the Triangle Multiplicative update. Unsurprisingly, this results in a large difference with the Baseline as these are the two main components contributing to the Pair representation, which we found to be critical to structure prediction performance (i.e., zeroing out Pair Representations results in large differences with the Baseline).

\section{Baseline Difference Versus Protein Length Plots}

\begin{figure}[h]
  \centering
  \begin{subfigure}[t]{0.32\linewidth}
    \centering
    \includegraphics[width=\linewidth]{final_images/delta_vs_length__ablate_msa_col.png}
    \caption{Skip MSA Col. Attention}
    \label{fig:del-msa-col-supp}
  \end{subfigure}\hfill
  \begin{subfigure}[t]{0.32\linewidth}
    \centering
    \includegraphics[width=\linewidth]{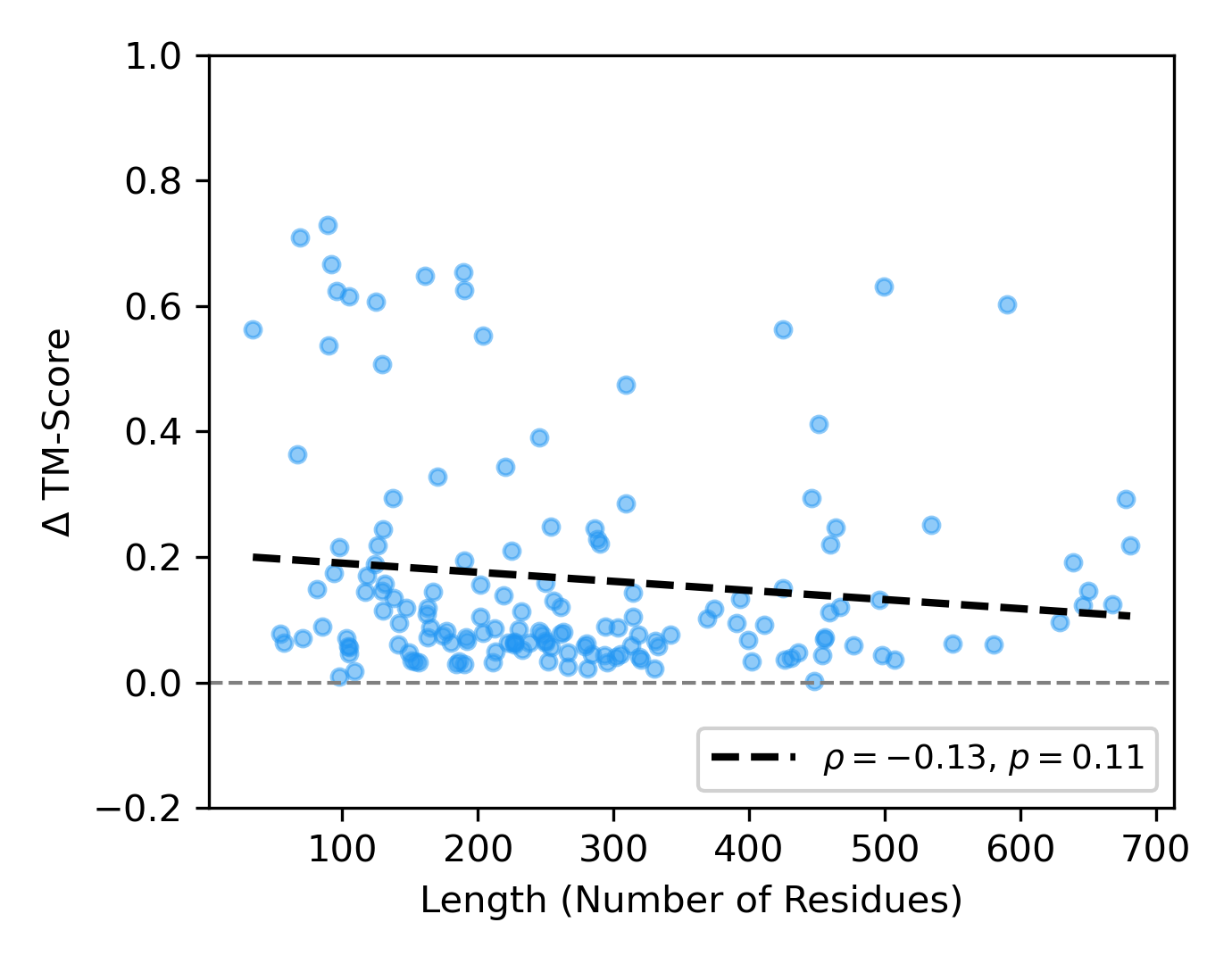}
    \caption{Only MSA Col. Attention}
    \label{fig:only-msa-col-supp}
  \end{subfigure}\hfill
  \begin{subfigure}[t]{0.32\linewidth}
    \centering
    \includegraphics[width=\linewidth]{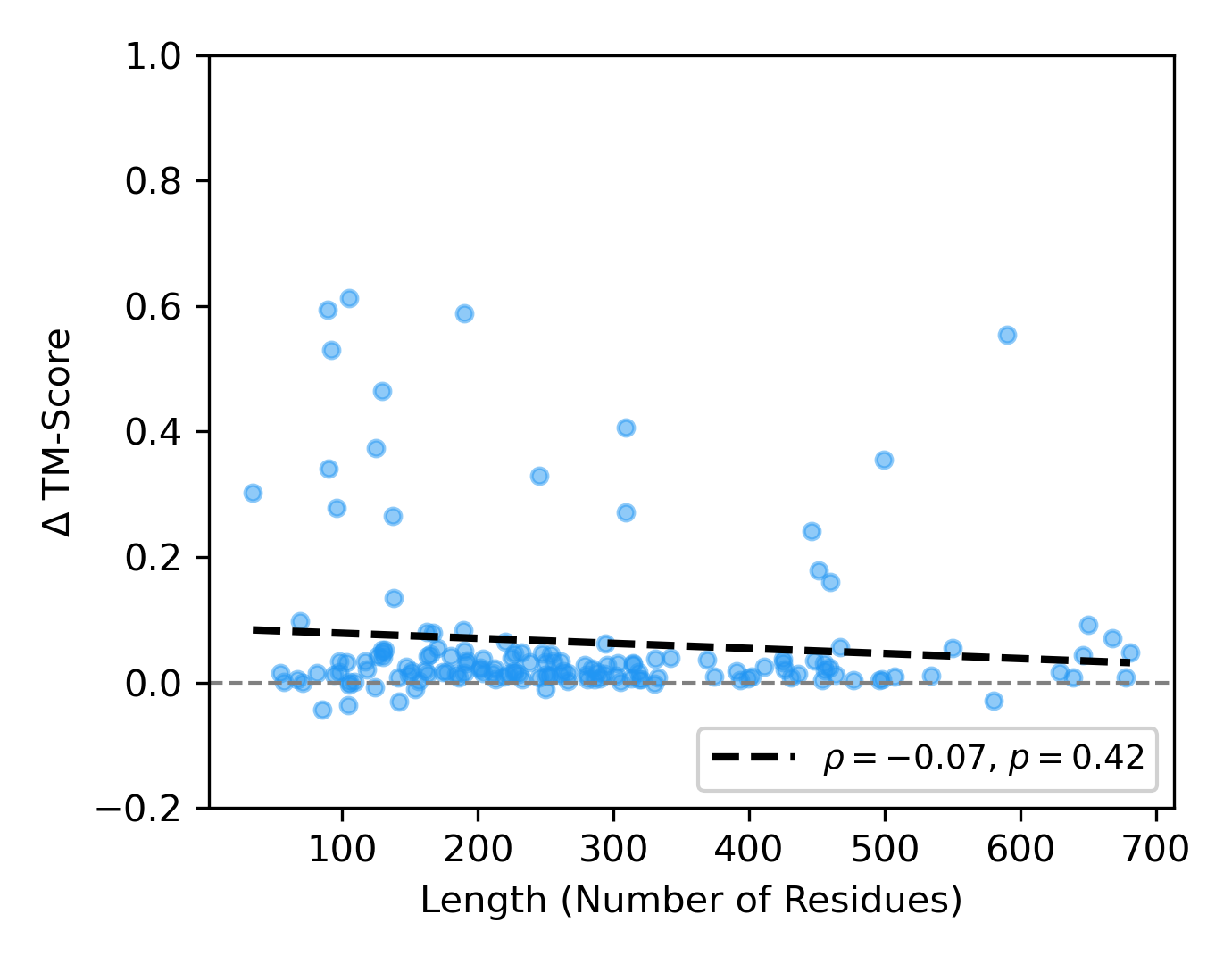}
    \caption{Skip MSA Row Attention}
    \label{fig:del-msa-row-supp}
  \end{subfigure}\\
    \begin{subfigure}[t]{0.32\linewidth}
    \centering
    \includegraphics[width=\linewidth]{final_images/delta_vs_length__ablate_tri_start_ablate_tri_end.png}
    \caption{Skip Triangle Attention}
    \label{fig:del-tri-supp}
  \end{subfigure}\hfill
      \begin{subfigure}[t]{0.32\linewidth}
    \centering
    \includegraphics[width=\linewidth]{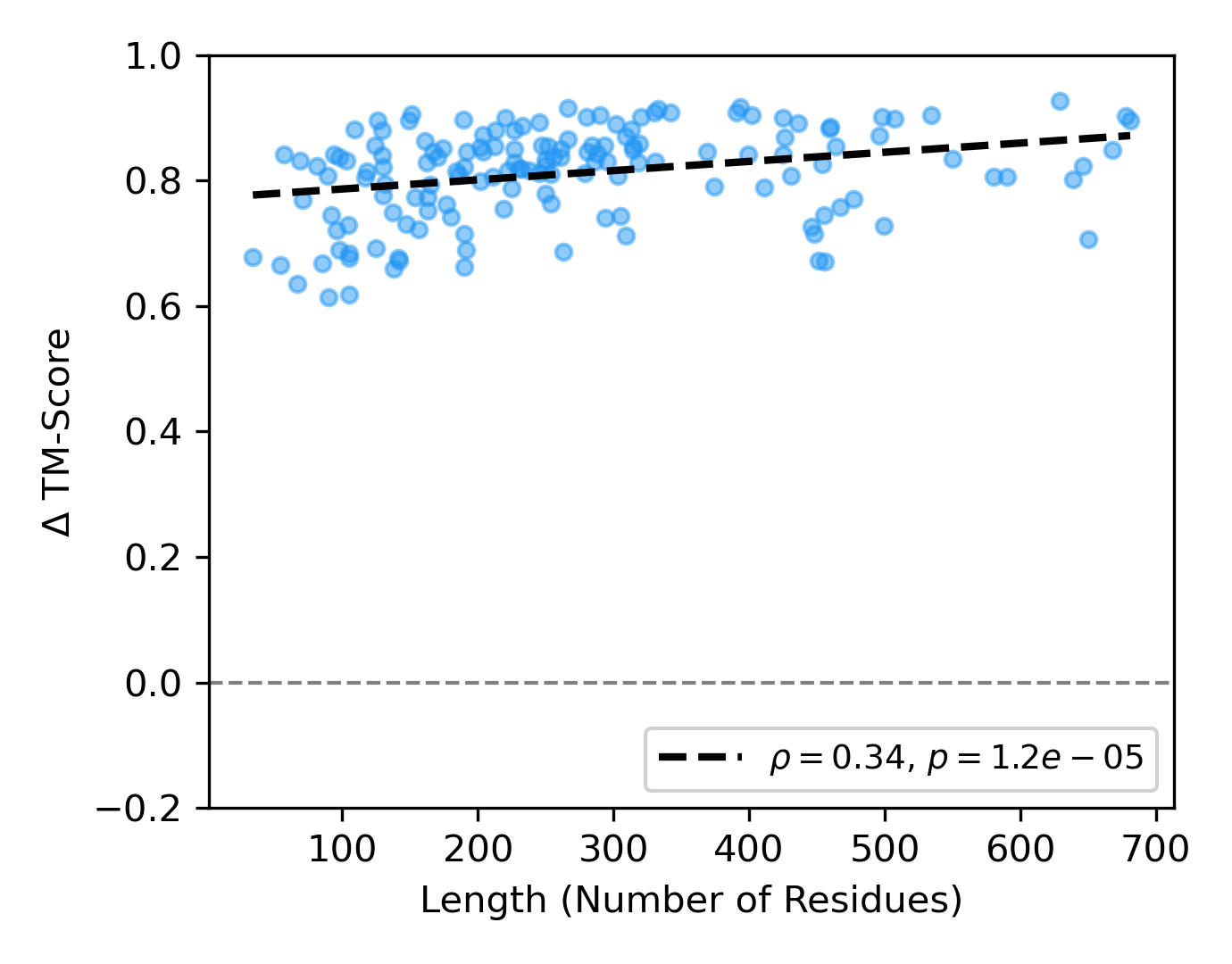} 
    \caption{Skip MSA Transition}
    \label{fig:del-msa-trans-supp}
  \end{subfigure}\hfill
    \begin{subfigure}[t]{0.32\linewidth}
    \centering
    \includegraphics[width=\linewidth]{final_images/delta_vs_length__zero_m_before_sm.png}
    \caption{Zero MSA Representation}
    \label{fig:zero-msa-supp}
  \end{subfigure}\\
      \begin{subfigure}[t]{0.32\linewidth}
    \centering
    \includegraphics[width=\linewidth]{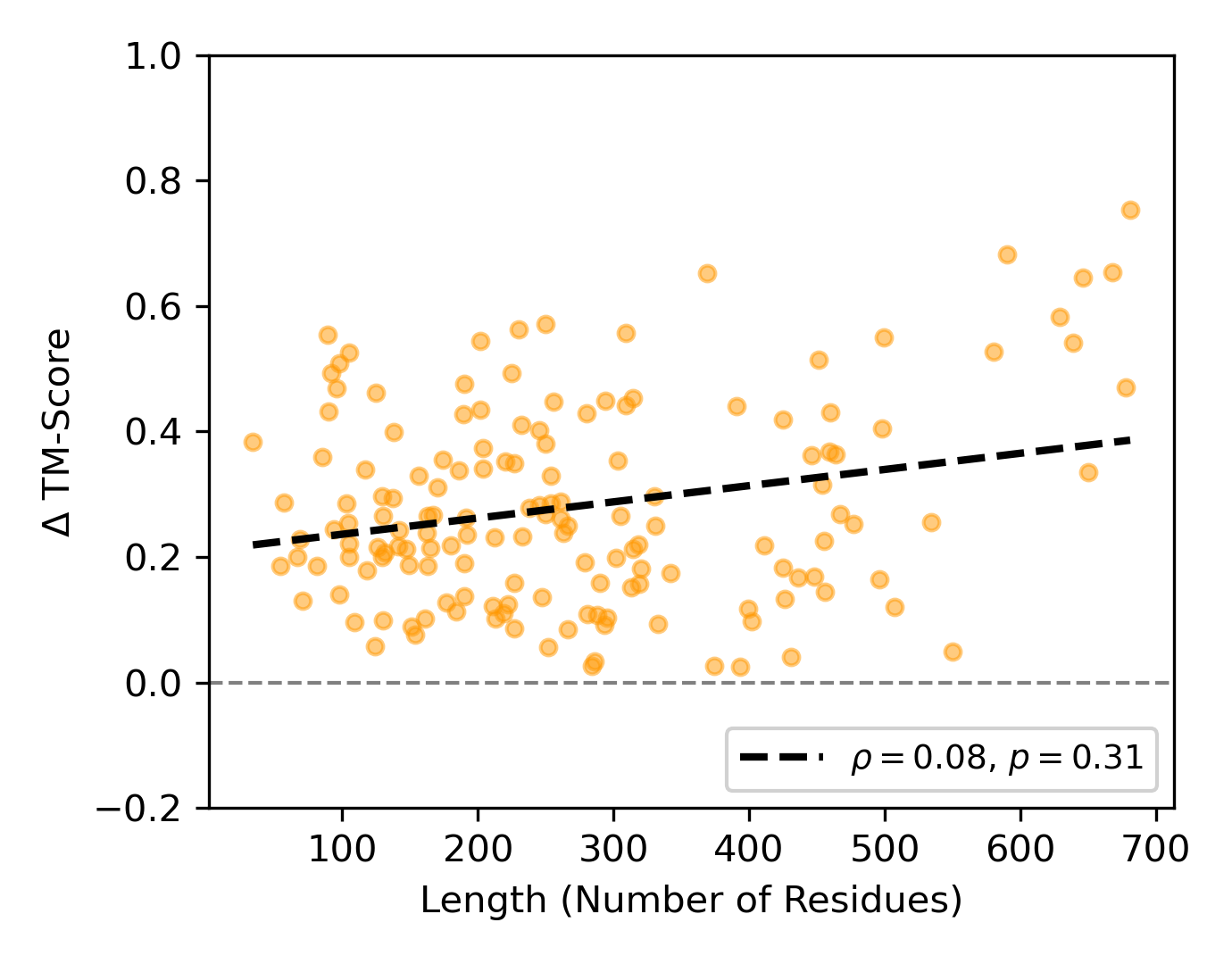}
    \caption{Skip Triangle Mult. Update}
    \label{fig:del-tmu-supp}
  \end{subfigure}\hfill
      \begin{subfigure}[t]{0.32\linewidth}
    \centering
    \includegraphics[width=\linewidth]{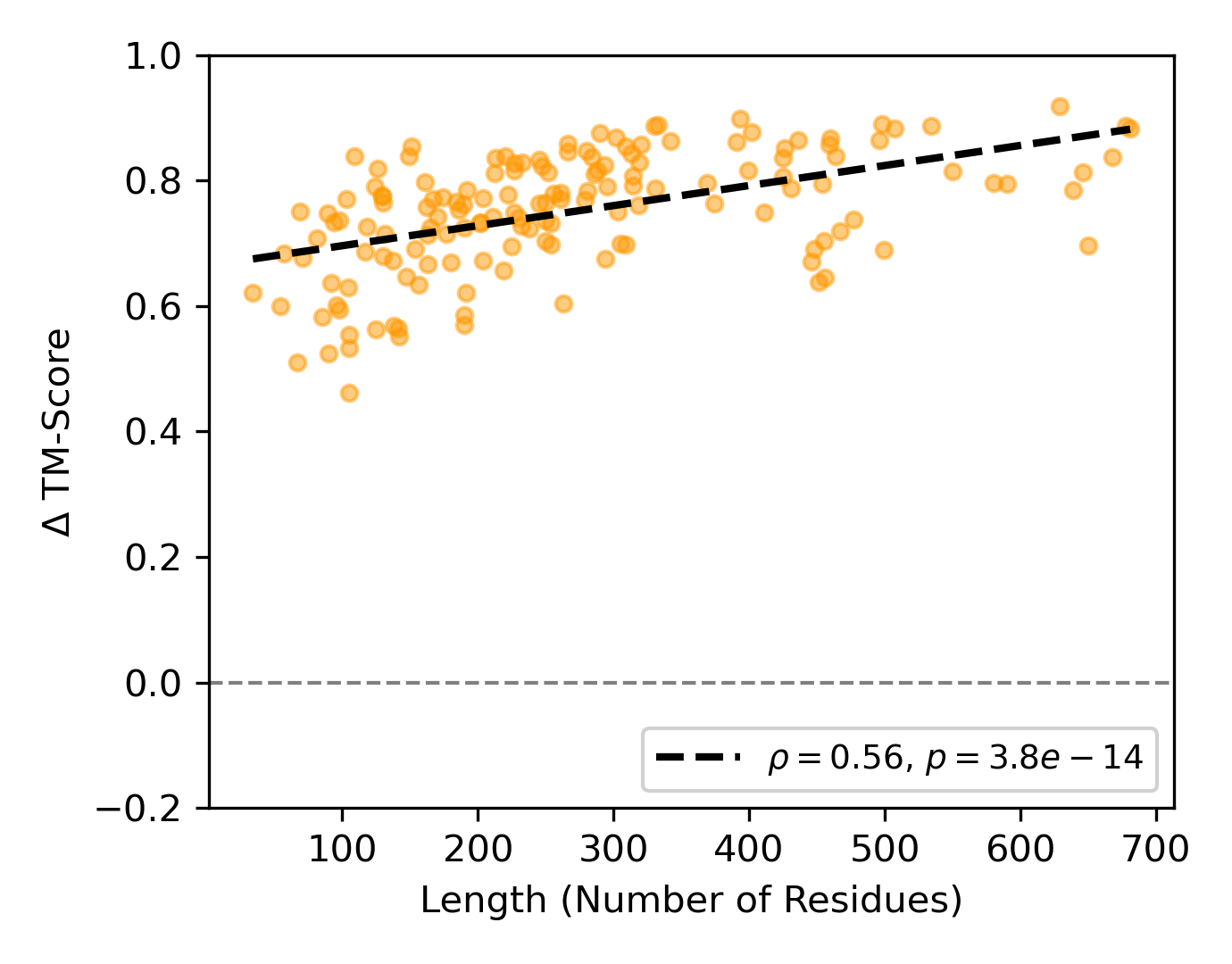}
    \caption{Skip Pair Transition}
    \label{fig:del-pair-trans-supp}
  \end{subfigure}\hfill
  \begin{subfigure}[t]{0.32\linewidth}
    \centering
    \includegraphics[width=\linewidth]{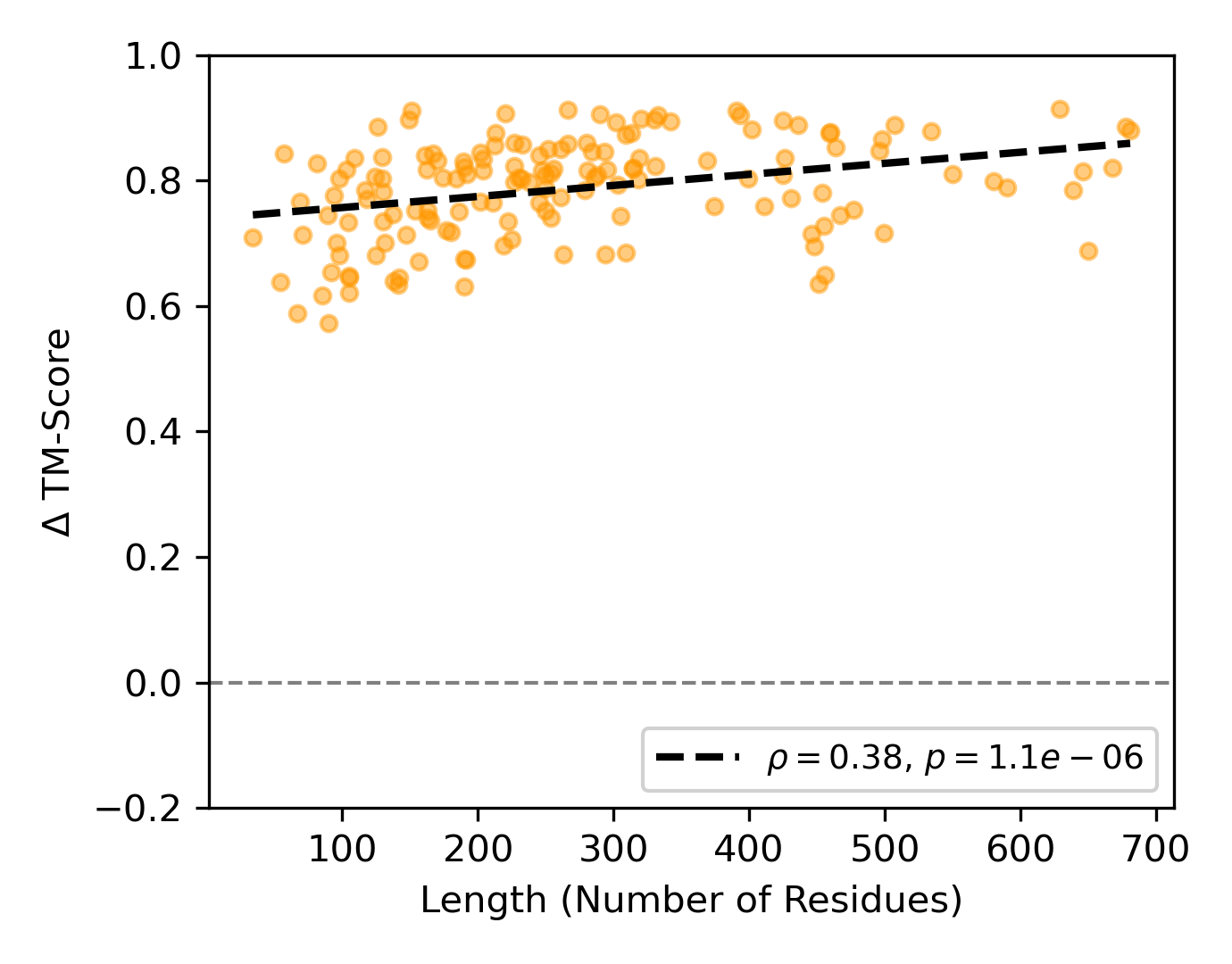}
    \caption{Zero Pair Representation}
    \label{fig:zero-pair-supp}
  \end{subfigure}\hfill
  \caption{\textbf{Comparison of Baseline Differences in TM-Scores ($\Delta$) vs. Protein Length.}
  Higher scores indicate more deviation from the Baseline. We show the best-fit line, the Spearman correlation coefficient, and its associated $p$-value. Each point represents the results for one protein.}
  \label{fig:scatter-plots-supp}
\end{figure}

In Fig.~\ref{fig:scatter-plots-supp}, we show the Baseline Difference versus Protein Length plots for all component studies carried out in our main paper. The correlation values for these experiments are included in Table~\ref{tab:correlation}.

\section{Replacing Representations with Noise}

\begin{figure}[t]
  \centering
  \includegraphics[width=0.7\linewidth]{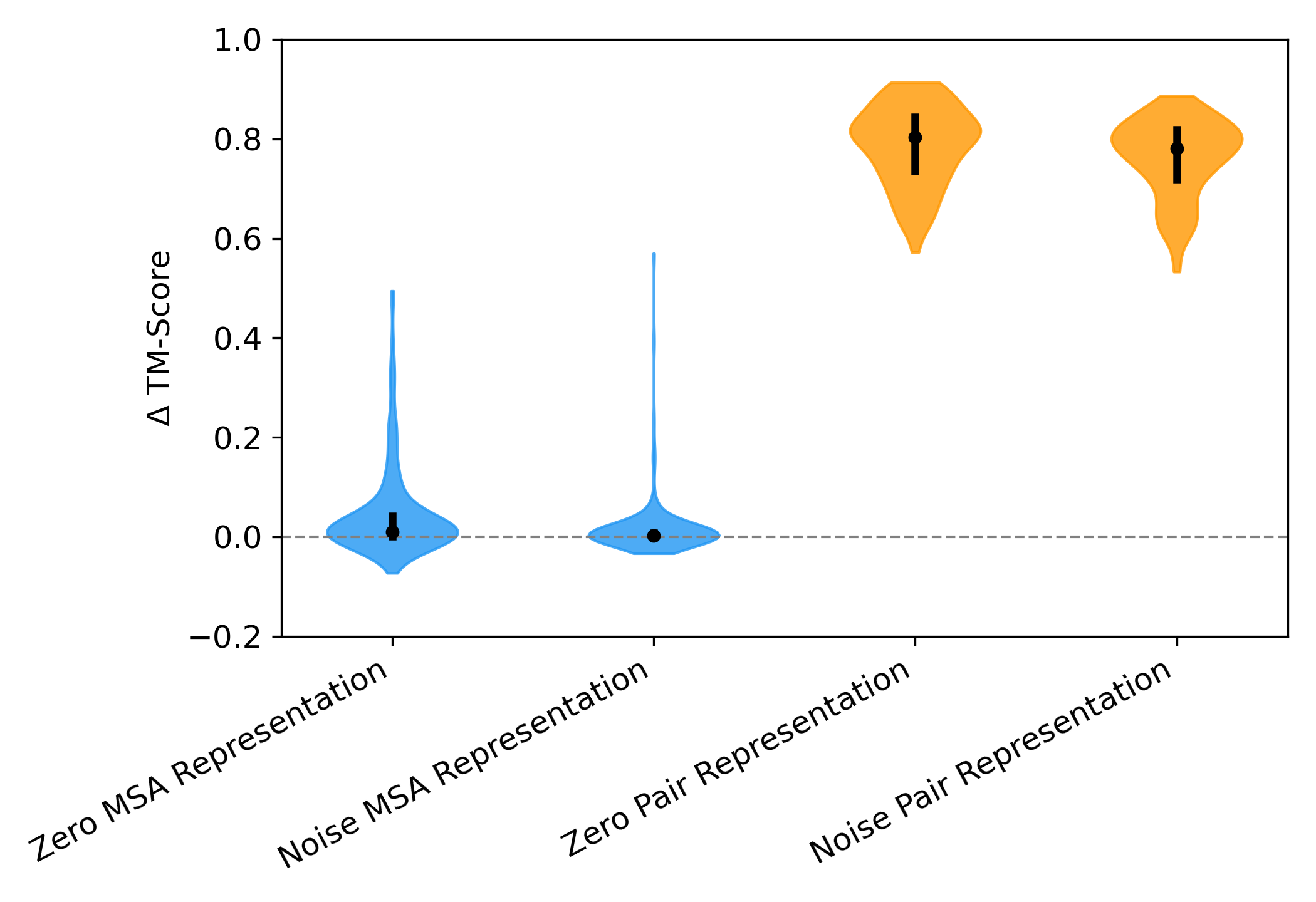}
  \caption{\textbf{Replacing MSA Representation and Pair Representation with Zeros vs. Noise.} We plot the differences ($\Delta$) between Baseline TM-Scores and each study. Higher scores indicate more deviation from the Baseline. Studies involving MSA and Pair Representations are in \color{blue}{blue} \color{black}{and} \color{orange}{orange}\color{black}{, respectively.}}
  \label{fig:noise}
\end{figure}

In our main experiments, we evaluated the impact of the MSA representation and the Pair representation on structure prediction by zeroing out their respective tensors before the Structure Module. Here, we instead ablate these representations by replacing them with random noise while preserving tensor scale. Specifically, for a representation tensor $\mathbf{X}$, we compute the per-channel mean $\bm{\mu}$ and standard deviation $\bm{\sigma}$ across the spatial dimensions (sequence or pair indices), and draw noise as $\bm{\mu} + \bm{\sigma} \odot \mathcal{N}\left(0,\mathbf{I}\right)$. We then replace $\mathbf{X}$ with this noise. The results, shown in Fig.~\ref{fig:noise}, mirror the zeroing experiments: replacing the MSA representation has minimal effect relative to the Baseline, while replacing the Pair representation leads to a large drop in TM-score.

\end{document}